\documentclass[11pt]{article}
\usepackage{latexsym}

\usepackage{amssymb}
\usepackage{amsmath}

 \usepackage{epsfig}
 \usepackage{graphicx}

\newcommand{\be}{\begin{equation}}
\newcommand{\ee}{\end{equation}}
\newcommand{\ba}{\begin{eqnarray}}
\newcommand{\ea}{\end{eqnarray}}
\newcommand{\bi}{\begin{itemize}}
\newcommand{\ei}{\end{itemize}}
\newcommand{\baa}{\begin{array}}
\newcommand{\eaa}{\end{array}}
\newcommand{\nn}{\nonumber \\}
\newcommand{\nr}[1]{(\ref{#1})}
\newcommand{\la}[1]{\label{#1}}

\newcommand{\rmi}[1]{{\mbox{\scriptsize #1}}}
\newcommand{\fr}[2]{{\frac{#1}{#2}\,}}
\newcommand{\fra}[2]{\textstyle{\frac{#1}{#2}\,}}  

\newcommand{\bfx}{{\bf x}}

\def\CL{{\cal L}}

\def\gsim{\raise0.3ex\hbox{$>$\kern-0.75em\raise-1.1ex\hbox{$\sim$}}}
\def\lsim{\raise0.3ex\hbox{$<$\kern-0.75em\raise-1.1ex\hbox{$\sim$}}}

\begin{document}

\begin{titlepage}
\begin{flushright}
HIP-2009-31/TH\\
December 2009\\
\end{flushright}
\begin{centering}
\vfill

{\Large{\bf Thermodynamics of a field theory with infrared fixed point from gauge/gravity
duality}}

\vspace{0.8cm}

\renewcommand{\thefootnote}{\fnsymbol{footnote}}

J. Alanen$^{\rm a,b}$\footnote{janne.alanen@helsinki.fi},
K. Kajantie$^{\rm a,b}$\footnote{keijo.kajantie@helsinki.fi}

\setcounter{footnote}{0}

\vspace{0.8cm}

{\em $^{\rm a}$%
Department of Physics, P.O.Box 64, FI-00014 University of Helsinki,
Finland\\}
{\em $^{\rm b}$%
Helsinki Institute of Physics, P.O.Box 64, FI-00014 University of
Helsinki, Finland\\}

\vspace*{0.8cm}

\end{centering}

\noindent
We use gauge/gravity duality to
study the thermodynamics of a field theory with asymptotic freedom in the
ultraviolet and a fixed point in the infrared. We find a high temperature
quark-gluon phase and a low $T$ conformal unparticle phase.
The phase transition between the phases is of first order or
continuous, depending on the ratio of the radii of asymptotic AdS$_5$ spaces
at $T=0$ and $T=\infty$. This is a prediction from a model of
gauge/gravity duality, not yet verified on the field theory side.

\vfill \noindent

%

\vspace*{1cm}

\noindent


\vfill

\end{titlepage}

\section{Introduction}
A gauge/gravity duality model for SU($N_c$) gauge theory thermodynamics
with a high temperature gluon plasma phase and a low $T$ glueball gas phase
and a first order transition in between has been presented in
\cite{kiri3,kiri4}. The model contains a 5-dimensional
metric ansatz and a scalar field with a bulk potential. A variant of
the model has been developed in \cite{aks}, in which the scalar field
potential as a starting point is replaced by the beta function of the
boundary field theory.

The purpose of this article is to apply the method in \cite{aks} to a
boundary field theory with the model beta function \cite{kiri1,kiri2}
\be
\beta(\lambda)=-\beta_0\lambda^2\left(1-{\lambda\over\lambda_*}\right),\qquad
\lambda=N_cg^2,
\la{betafn}
\ee
where $\beta_0>0$ is a parameter and $\lambda=\lambda_*$ is an infrared (IR)
fixed point. What is the finite temperature phase structure of this theory?
At high temperatures
one expects the existence of a "quark-gluon plasma" phase\footnote{
Realistic theories of the type \nr{betafn} have fermions and
we assume that we can apply the model in \cite{aks} to them.},
conformal at $T\to\infty$.
Due to the IR fixed point one knows that there are no massive
glueball states in the theory. Thus there is no low $T$ glueball phase but,
instead, a conformal "unparticle" phase of massless particles. We shall
show that, indeed, this picture follows naturally from the model in
\cite{kiri3,aks}. Furthermore, the transition is of first order for
$\beta_0\lambda_*>6.58$ and a crossover for smaller values.

More formally, at $T=0$ and $T=\infty$ we have two pure AdS$_5$ spaces
with radii $\CL_\rmi{IR}$ and $\CL_\rmi{UV}$ and the transition we study is
driven by varying $T$. As we shall see,
the parameter $\beta_0$ is essentially the ratio of the radii:
\be
\beta_0\lambda_*=\fra92\log{\CL_\rmi{UV}\over\CL_\rmi{IR}}.
\la{betaL}
\ee

Beta functions of the type in Eq.\nr{betafn} but somewhat more complicated
have recently become a central topic of research in connection of
technicolor theories
\cite{sanninotuominen,ryttovsannino,jarvinen,antipintuominen}
and lattice Monte Carlo studies thereof
\cite{appelquist,catterall,degrand,hietanen,deldebbio,fodor,kogut}.
The results in this article should set the stage for the study of the
thermodynamics of these more general theories. Is it possible to use
lattice Monte Carlo
to search for the two-phase structure predicted here and to determine the
order of the transition?

\section{The gravity sector of the model}
We summarize first the gravity sector equations of the model in \cite{kiri3,aks}.
The model starts from a metric ansatz
\be
ds^2=b^2(z)\left[-f(z)dt^2+d\bfx^2+{dz^2\over f(z)}\right]
\la{ansatz}
\ee
plus a scalar field $\phi(z)=\log\lambda(z)$, where the four functions
$b(z), f(z)$ in the metric, the scalar field $\phi(z)$ and the
potential $V(\phi(z))$ in the gravity action (in standard notation)
\be
S={1\over16\pi G_5}\int d^5x\,\sqrt{-g}\left[R-\fra43(\partial_\mu\phi)^2+V(\phi)\right]
\la{Eframeaction}
\ee
are determined from the four equations ($\dot b\equiv b'(z)$, etc.)
\ba
&&6{\dot b^2\over b^2}+3{\ddot b\over b}+3{\dot b\over b}{\dot f\over f}={b^2\over f}V(\phi),\label{eq1}\\
&& 6{\dot b^2\over b^2}-3{\ddot b\over b}={\fra43} \dot\phi^2,\label{eq2}\\
&&{\ddot f\over \dot f}+3{\dot b\over b}=0,\label{eq3}\\
&&\beta(\lambda)=b{d\lambda\over db},\quad \lambda(z)  = e^{\phi(z)}\sim  g^2N_c.
\label{crucial}
\ea

First, from \nr{crucial} it follows that, for the beta function \nr{betafn},
\be
\log{b(\lambda)\over b_0}={1\over\hat\beta_0}
\left[{1\over\hat\lambda}-\log{\hat\lambda\over 1-\hat\lambda}\right]
\equiv Q+{1\over\hat\beta_0}\log(\hat\beta_0 Q-1),
\la{bQ}
\ee
where we have scaled (hats will be removed after Eq.~\nr{Lambdaz})
\be
\hat\lambda = {\lambda\over\lambda_*},\qquad \hat\beta_0=\lambda_*\beta_0
\ee
and where we have introduced \cite{aks}
\be
Q\equiv {1\over\hat\beta_0\hat\lambda}.
\ee
Further, to solve the second equation one introduces
\be
W= -\dot b/b^2,
\la{defW}
\ee
and solves
\ba
W(\lambda)&=&W(0)\exp\left(-\fra49\int_0^\lambda d\bar\lambda{\beta(\bar\lambda)
\over\bar\lambda^2}\right)=W(0)\,e^{\fr49\hat\beta_0(\hat\lambda-\fr12\hat\lambda^2)}\nn
&=&{1\over\CL}e^{{2\over9}\hat\beta_0}e^{-{2\over9}\hat\beta_0(1-\hat\lambda)^2}
={1\over\CL}e^{{4\over9Q}-{2\over9\hat\beta_0 Q^2}}.\la{W}
\ea
Given $W$ one can immediately write down \cite{kiri1,kiri2}
the scalar field potential $V(\phi)$
corresponding to $f(z)=1$ in \nr{eq1} (from now on hats are removed):
\be
V(\lambda=e^\phi) ={12\over\CL^2}e^{{4\over9}\beta_0}\,e^{-{4\over9}\beta_0(1-\lambda)^2}
\left[1-\fra19\beta_0^2\lambda^2(1-\lambda)^2\right].
\la{Vvac}
\ee
In the UV ($\lambda=0$) and IR ($\lambda=1$) limits we have two AdS$_5$ spaces with
radii $\CL=\CL_\rmi{UV}$ and $\CL_\rmi{IR}=\CL_\rmi{UV}e^{-2\beta_0/9}$. Eq. \nr{betaL}, giving
the parameter $\beta_0$ as essentially their ratio, follows immediately from this.

When $f$ is the solution of \nr{eq3}, \nr{eq1} gives the potential
\be
V(\lambda)=12fW^2\left[1-\left({\beta\over3\lambda}\right)^2\right]-3{\dot f\over b}W.
\la{fV}
\ee

From Eq.~\nr{defW} one further obtains,
defining the energy scale $\Lambda=b_0/\CL$,
\be
\Lambda z=\int_Q^\infty {dQ\over(\beta_0Q-1)^{1/\beta_0+1}}\beta_0Q\,e^{-Q-{4\over9Q}+
{2\over9\beta_0Q^2}}.
\la{Lambdaz}
\ee
Finally, from Eq.~\nr{eq3} we solve
\be
f(z)=1-\int_0^z{d\bar z\over b^3(\bar z)}/\int_0^{z_h}{d\bar z\over b^3(\bar z)},
\ee
where two constants of integration are fixed by $f(0)=1$ and $f(z_h)=0$.

Thermodynamics now is obtained from the key relations
\be
4\pi T=-f'(z_h),\qquad S={A\over4G_5} = {b^3\over4G_5} V_3.
\ee
For the temperature we thus have
\be
{1\over4\pi T}= b^3\int_0^z{dz\over b^3}=b^3\int_0^\lambda {d\lambda\over -\beta b^4W}=
b^3\int_Q^\infty{dQ\over 1-1/(\beta_0 Q)}{1\over b^4W}
\ee
or
\be
{\Lambda\over4\pi T}=e^{3Q}(\beta_0Q-1)^{3/\beta_0}
\int_Q^\infty {dQ\over (\beta_0Q-1)^{4/\beta_0+1}}\,\beta_0Q\,e^{-4Q-{4\over9Q}
+{2\over9\beta_0Q^2}}.
\la{invt}
\ee
The derivative $dT/dQ$ will be very important in what follows and this is
computed to be
\be
{dT\over T dQ}={\beta_0 Q\over \beta_0 Q-1}\left({4\pi T\over bW}-3\right).
\ee
Full thermodynamics is then given by
\ba
s(T)&=&{1\over 4G_5}b^3(Q(T)),\\
p(T)&=&\int^T dT s(T)={1\over4G_5}\int_0^{Q(T)}dQ\,{dT\over dQ}b^3(Q),\la{pee2}\\
\epsilon(T)&=&
{3\over4G_5}\int_0^{Q(T)}dQ\,\,T(Q)\,b^3(Q){d\log b\over dQ},\la{epsil2}\\
\epsilon(T)-3p(T)&=&
{3\over4G_5}\int_0^{Q(T)}dQ\,\left(T{d\log b\over dQ}-{dT\over dQ}\right)\,b^3(Q),\\
c_s^2&=&{1\over3T}\,{dT\over dQ}{dQ\over d\log b}=\fr13\left(4{\pi T\over bW}-3\right)={s\over C_V},
\la{cs2}
\ea
where $b(Q)$ is given by \nr{bQ} and $T(Q)$ by \nr{invt}.

\section{Analytic approximations in the ultraviolet and infrared}
Final results will have to be computed numerically, but it is very useful to have
analytic approximations in the ends of the phase space, UV ($Q\to\infty$)
and IR ($Q\to1/\beta_0$).

The leading UV terms are very simple:
\be
{b\over b_0}=e^Q (\beta_0 Q)^{1\over\beta_0}={1\over\Lambda z}
\equiv{\CL\over b_0z}={\pi T\over\Lambda};
\la{UV}
\ee
further corrections can be easily evaluated. The UV limit of bulk thermodynamic
quantities is easiest to obtain from
\be
{s\over T^3} ={\CL^3\over 4G_5}\,
{(b(Q\to\infty)/b_0)^3\over(T(Q\to\infty)/\Lambda)^3}=
{\CL^3\over 4G_5}\,\pi^3
\ee
since the UV limit here is nothing but $1/(z_hT)^3=\pi^3$. Equivalently,
\be
{p\over T^4}={\CL^3\over4G_5}\,{\pi^3\over4},\qquad T\to\infty.
\la{pT4as}
\ee
Normalising to ideal gas pressure $p=g_\rmi{eff}\pi^2T^4/90$ one has
\be
{\CL^3\over 4G_5} ={2\over45\pi}g_\rmi{eff}.
\la{norm}
\ee

The conformal IR region is
the region near $\beta_0Q-1=0$ where the powerlike behavior of $b(Q)$ drives
it to zero.
For $b$ and $W$ one can trivially expand:
\ba
b&=&b_0e^{1\over\beta_0}(\beta_0Q-1)^{1\over\beta_0}
\left[1+{1\over\beta_0}(\beta_0Q-1)+{1\over2\beta_0^2}(\beta_0Q-1)^2+...\right],
\la{bconf}\\
W&=&{1\over\CL}e^{\fr29\beta_0}\left[1-{2\beta_0\over9}(\beta_0Q-1)^2+
{\cal O}(\beta_0Q-1)^3\right].
\ea
For $\Lambda z$ and $T/\Lambda$ the situation is more subtle: to ensure convergence
of the integrals \nr{Lambdaz} and \nr{invt} at the lower limit
one must partially integrate in the power of $\beta_0Q-1$ to decrease the exponent.
One partial integration gives the leading power.
For $\Lambda z$ one has
\be
\Lambda z=[e^{-{1\over\beta_0}-\fr29\beta_0}(\beta_0Q-1)^{-{1\over\beta_0}}+I_z(\beta_0)]
[1+{\cal O}(\beta_0Q-1)],
\la{zleading}
\ee
where
\be
I_z(\beta_0)=\int_{1/\beta_0}^\infty{dQ\over(\beta_0Q-1)^{1/\beta_0}}{d\over dQ}\left[\beta_0Q\,
e^{-Q-{4\over9Q}+{2\over9\beta_0Q^2}}\right]
\ee
is an integral arising from partial integration. This converges for $\beta_0>1$.
If $\beta_0<1$ more partial integrations are needed to obtain finite subleading terms.
Anyway the leading term \nr{zleading} implies that the fields always extend to $z=\infty$
in the IR limit.

For $T$ the leading term is
\be
{T(\beta_0Q\to 1)\over\Lambda} =
{1\over\pi}e^{{1\over\beta_0}+{2\beta_0\over9}}\left(\beta_0Q-1\right)^{{1\over\beta_0}}
\left[1-e^{{4\over\beta_0}+{2\beta_0\over9}}I_T(\beta_0)(\beta_0Q-1)^{4\over\beta_0}
\right],
\la{tvanish}
\ee
where
\be
I_T(\beta_0)=\int_{1/\beta_0}^\infty{dQ\over(\beta_0Q-1)^{4/\beta_0}}{d\over dQ}\left[\beta_0Q\,
e^{-4Q-{4\over9Q}+{2\over9\beta_0Q^2}}\right], \qquad \beta_0>4,
\la{ip}
\ee
converges if $\beta_0>4$.
Using this in \nr{cs2} one obtains for the sound velocity in the conformal region
\be
c_s^2=\fr13\left[1-4e^{{4\over\beta_0}+{2\beta_0\over9}}I_T(\beta_0)(\beta_0Q-1)^{4\over\beta_0}
\right].
\la{csconf}
\ee

For the pressure the next-to-leading order expansion in
the IR conformal domain is (see Figs.~\ref{fig3} and \ref{fig4})
\be
{p\over T^4}={\CL^3\over 4G_5} {\pi^3\over4} e^{-{2\beta_0\over3}}
\left[1+\fra32 e^{{4\over\beta_0}+{2\beta_0\over9}}I_T(\beta_0)(\beta_0Q-1)^{4\over\beta_0}
+{\cal O}(\beta_0Q-1)\right].
\la{pT4conf}
\ee
The overall magnitude here is the same as at large $T$ but suppressed by the
factor $e^{-{2\beta_0\over3}}$.


\begin{figure}[!tb]
\begin{center}

\includegraphics[width=0.6\textwidth]{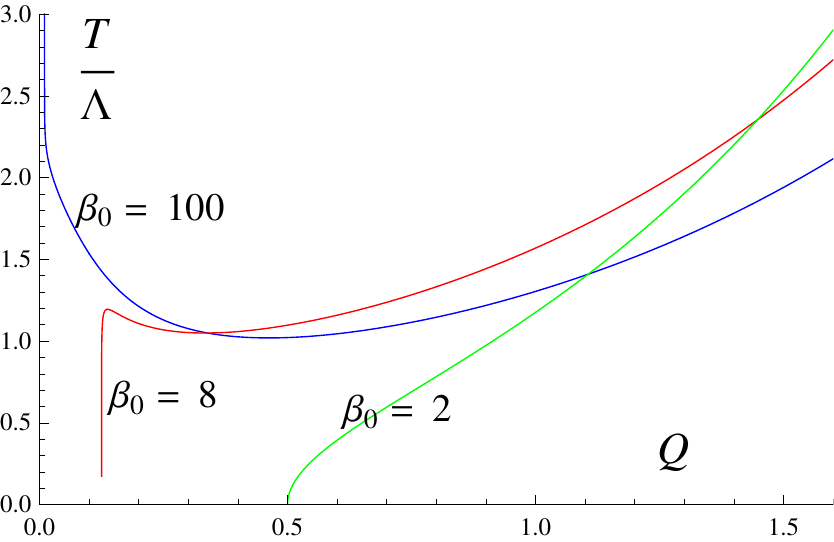}
\end{center}

\caption{\small Temperature vs. $Q=1/(\beta_0\lambda)$ as computed from \nr{invt}.
The minimum disappears for
$\beta_0<6.58$. Always $T(Q=1/\beta_0)=0$, see Eq.\nr{tvanish}.
}\la{fig1}
\end{figure}

\begin{figure}[!tb]
\begin{center}

\includegraphics[width=0.46\textwidth]{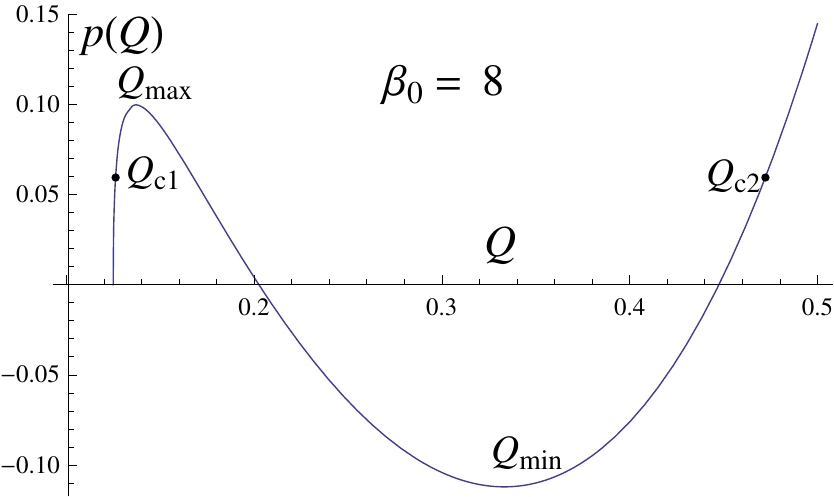}\hfill
\includegraphics[width=0.46\textwidth]{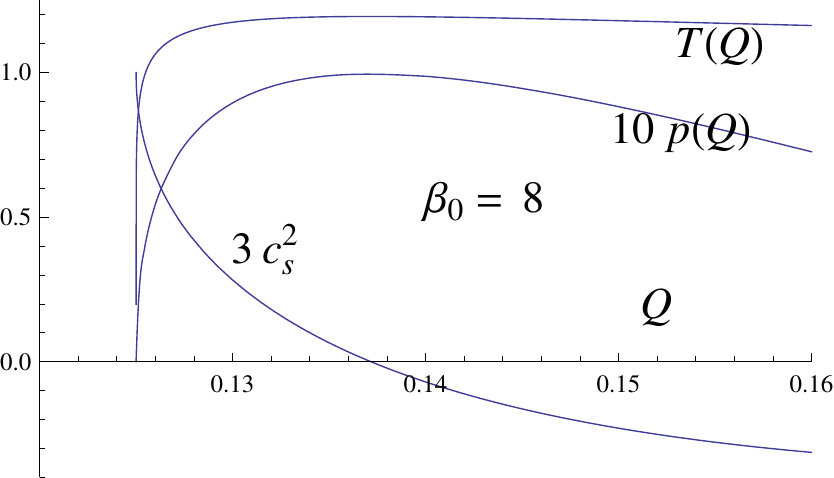}
\end{center}

\caption{\small (Left panel) The pressure for $\beta_0=8$.
Note the positive peak at $Q=0.137$, caused via Eq.~\nr{pee2} by
the increase of $T(Q)$ starting from the IR limit $1/\beta_0=0.125$.
(Right panel) Field configurations in the conformal region for $\beta_0=8$.
Note that the sound velocity vanishes at the same point
$Q=Q_\rmi{max}=0.137$ where $p'(Q)$ and $T'(Q)$ vanish.
}\la{fig2}
\end{figure}

\begin{figure}[!tb]
\begin{center}

\includegraphics[width=0.6\textwidth]{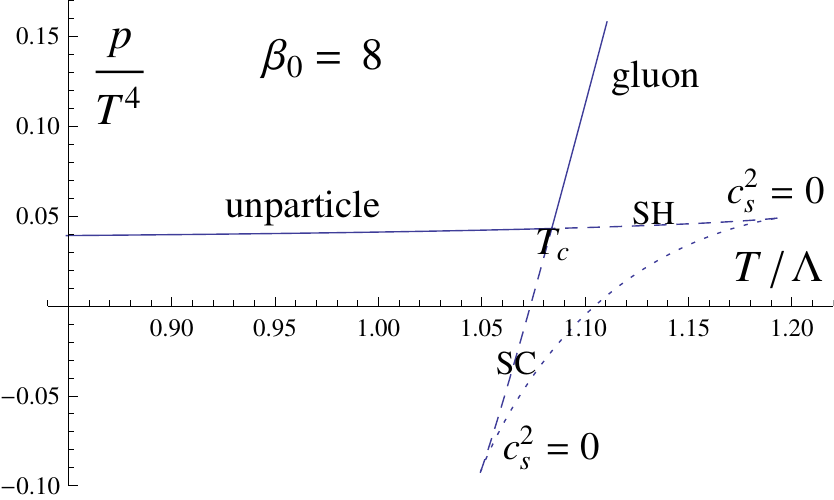}
\end{center}

\caption{\small Plot of $p/T^4$ vs $T/\Lambda$ in the critical region for a non-monotonic
$T(Q)$. There is a stable high $T$ gluon
plasma phase with a metastable supercooled (SC) branch and a stable low $T$ unparticle
phase with a metastable superheated (SH) branch. The metastable branches terminate at the
points where sound velocity squared becomes negative;
between these points the system is mechanically
unstable. There is a first order phase transition at $T_c=1.084\Lambda$.
The unparticle phase
extends to $T=0$ to the value $\pi^3\,e^{-16/3}/4=0.037$, the gluon plasma phase at
$T\gg T_c$ to $\pi^3/4=7.75$.
For normalisation to ideal gas, multiply by the factor in Eq.~\nr{norm}.
}\la{fig3}
\end{figure}

\section{The two phases}
To clarify the structure of the solution, the behavior of $T(Q)$ for a few values of $\beta_0$
is shown in Fig.~\ref{fig1}. For any value of $\beta_0$, $T$ vanishes
$T(Q)/\Lambda\sim(\beta_0Q-1)^{1/\beta_0}$
at the infrared endpoint, see Eq.~\nr{tvanish}. For
$Q\gg1/\beta_0$ the limit is $T(Q)/\Lambda\sim e^Q$, Eq.~\nr{UV}.
In between, $\beta_0=6.58$ separates two qualitatively different behaviors.
For $\beta_0<6.58$ $T(Q)$ grows monotonically with $Q$, $dT/dQ>0$ always. For
$\beta_0>6.58$, $T(Q)$ has a maximum and minimum, i.e., two zeroes of $dT/dQ$ and a
range where $dT/dQ<0$, $c_s^2<0$ so that the system is mechanically unstable.

To see concretely what this implies for physics, consider separately the two
cases $\beta_0=8$ with two zeroes of $dT/dQ$ and $\beta_0=2$ with $dT/dQ>0$.

Fig.~\ref{fig2} shows for $\beta_0=8$
a plot of $p(Q)$ and a blow-up of the far infrared region. The two zeroes of $T'(Q)$
correspond to the two zeroes of $p'(Q)$, a maximum and a minimum of $p(Q)$.
The physics is seen from the plot of pressure vs. $T$ in Fig.~\ref{fig3}. The
correspondence between Figs. \ref{fig2} and \ref{fig3} is as follows:
\bi
\item The range $1/\beta_0<Q$ up to the first maximum $Q_\rmi{max}$
of $p(Q)$, at which $c_s^2=0$,
corresponds to the unparticle phase in Fig. \ref{fig3}.
Only the part of this up to $Q_{c1}$ is thermodynamically stable,
the rest is a metastable superheated branch.
\item The interval $Q_\rmi{max}<Q<Q_\rmi{min}$ in which $p(Q)$ decreases in
Fig. \ref{fig2} corresponds to
the unstable phase with $c_s^2<0$ in Fig. \ref{fig3}.
\item The interval of $Q>Q_\rmi{min}$ in which $p(Q)$ again
increases in Fig. \ref{fig2} corresponds
to the quark-gluon plasma phase; only the part of this above $Q_{c2}$ is thermodynamically stable,
the rest is a metastable supercooled branch.
\item The phase transition takes place at $T_c=1.084\Lambda$. The latent heat is
numerically
\be
{\Delta\epsilon\over T_c^4}={\CL^3\over 4G_5}\,4.55,
\ee
where \nr{norm} gives the normalisation. A fit to nearby values is
$L/T_c^4=3.86 (\beta_0-6.58)^{0.427}$.
\ei

\begin{figure}[!b]
\begin{center}

\includegraphics[width=0.6\textwidth]{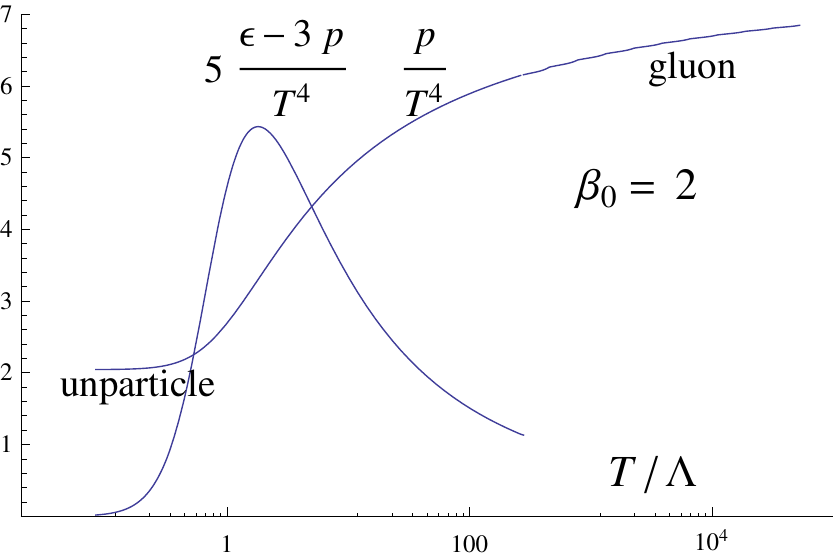}
\end{center}

\caption{\small Plot of $p/T^4$ and of the interaction measure
$(\epsilon-3p)/T^4$ (multiplied by 5)
vs $T/\Lambda$ for $\beta_0=2$ with a monotonic
$T(Q)$. The two phases are continuously connected. The unparticle phase
extends to $T=0$ to the value $\pi^3\,e^{-4/3}/4=2.043$, the gluon plasma phase at
$T\gg \Lambda$ to $\pi^3/4=7.75$.
For normalisation to ideal gas, multiply by the factor in Eq.~\nr{norm}.
}\la{fig4}
\end{figure}

If one now decreases $\beta_0$, the first order transition in Fig. \ref{fig3} becomes
weaker and terminates at $\beta_0=6.582$ at $T=0.991\Lambda$ and $p/T^4=0.134$ (which decreases
to $0.0963$ at $T=0$ for this $\beta_0$). Below that the transition is continuous.
The situation for $\beta_0=2$ is shown in Fig.~\ref{fig4}. Since $dT/dQ>0$ there is no range
with negative $c_s^2$ and the unstable phase in Fig. \ref{fig3} disappears. Related to
this, there is no discontinuity in $p'(T)$, the two phases are continuously connected.
Fig.~\ref{fig4} also shows the interaction measure
\be
{\epsilon-3p\over T^4}=T{\partial\over\partial T}{p\over T^4}.
\ee
Its maximum at $T=1.81\Lambda$ could be used as an estimate of the crossover
temperature at this $\beta_0$.

\section{Conclusions}
We have, in this article, studied the thermodynamics of a field theory with the
beta function \nr{betafn} with an infrared fixed point using a
gauge/gravity duality model \cite{kiri3,kiri4,aks}. In the two limits
$T=0$ and $T=\infty$ we have conformal theories with pure AdS$_5$ spaces as
gravity duals. The crucial parameter $\beta_0$ of this model is essentially
the ratio of the radii of these spaces, Eq. \nr{betaL}. Varying $T$ drives
a phase transition which in this model is of first order
for $\CL_\rmi{UV}/\CL_\rmi{IR}>4.32$, continuous otherwise.

The computations here used Eqs.\nr{eq2}-\nr{crucial} to determine the metric
functions, scalar field, and the phase structure \cite{aks}. However, we have also verified
that the same phase structure is obtained by integrating numerically the Einstein
equations \nr{eq1}-\nr{eq3} with the scalar potential in \nr{Vvac} and by computing
the equation of state as in \cite{kiri4}. Quantitative details change somewhat,
for example, the critical value of $\beta_0\lambda_*$ is 4.85 instead of 6.58 above.

It has been a very challenging task to identify four dimensional gauge
field theories with beta functions of the type \nr{betafn}. These need
fermions in special representations \cite{sanninotuominen} and the
proof requires extensive numerical effort
\cite{appelquist,catterall,degrand,hietanen,deldebbio,fodor,kogut}. It is clear that
the finite temperature properties of these theories will be studied
but the path to clarification of the phase structure will be long.

Theoretically it will be interesting to study the finite $T$ phase
structure of theories with a walking technicolor beta function, for
which the IR fixed point is only approached. The theory then
is confining and there is no conformal unparticle phase at $T=0$.
The model in this article can be used for that purpose.

\vspace{1cm}
{\it Acknowledgements}.
We thank K. Rummukainen, V. Suur-Uski and K. Tuominen  for discussions
and F. Nitti for the well documented numerical code used in \cite{kiri4}.
JA thanks the Magnus Ehrnrooth foundation for financial support.

%
%

%
%

\end{document}